\RequirePackage{rotating}
\documentclass[a4paper,fleqn,usenatbib]{mnras}

\usepackage{newtxtext,newtxmath}
\usepackage[T1]{fontenc}
\usepackage{ae,aecompl}
\usepackage{graphicx}	% Including figure files
\usepackage{amsmath}	% Advanced maths commands
\usepackage{amssymb}	% Extra maths symbols

\usepackage{rotating}
\usepackage{threeparttable}
\usepackage[font=small,format=plain,labelfont=bf,up,labelsep = period,singlelinecheck = false, justification=justified]{caption}
\usepackage{multirow}

\usepackage{color}

\def\aap{A\&A}
\def\apj{ApJ}
\def\mnras{MNRAS}
\def\apjs{ApJS}

\title[State Transitions in Ultracompact Neutron Star LMXBs: towards the Low Luminosity Limit]{State Transitions in Ultracompact Neutron Star LMXBs: towards the Low Luminosity Limit}

\author[Jie Lin and Wenfei Yu]{Jie Lin$^{1,2}$ and Wenfei Yu$^{1}$\thanks{E-mail:wenfei@shao.ac.cn}\\
$^{1}$Key Laboratory for Research in Galaxies and Cosmology and Shanghai Astronomical Observatory, Chinese Academy of Sciences,\\ 80 Nandan Road, Shanghai 200030,China;\\
$^{2}$University of Chinese Academy of Sciences, 19 A Yuquan Road, Beijing 100049, China;
}

\date{Accepted . Received ; in original form }
\pubyear{2017}

\begin{document}
\label{firstpage}
\pagerange{\pageref{firstpage}--\pageref{lastpage}}
\maketitle

% Abstract of the paper
\begin{abstract}
Luminosity of X-ray spectral state transitions in black hole and neutron star X-ray binaries can put constraint on the critical mass accretion rate between accretion regimes. Previous studies indicate that the hard-to-soft spectral state transitions in some ultracompact neutron star LMXBs have the lowest luminosity.  With X-ray monitoring observations in the past decade, we were able to identify state transitions towards the lowest luminosity limit in 4U 0614+091, 2S 0918-549 and 4U 1246-588. By analysing corresponding X-ray pointed observations with the Swift/XRT and the RXTE/PCA, we found no hysteresis of state transitions in these sources, and determined the critical mass accretion rate in the range of 0.002 -- 0.04 $\dot{\rm M}_{\rm Edd}$ and 0.003 -- 0.05 $\dot{\rm M}_{\rm Edd}$ for the hard-to-soft and the soft-to-hard transition, respectively, by assuming a neutron star mass of 1.4 solar masses. This range is comparable to the lowest transition luminosity measured in black hole X-ray binaries, indicating the critical mass accretion rate is not affected by the nature of the surface of the compact stars. Our result does not support the Advection-Dominated Accretion Flow (ADAF) model which predicts that the critical mass accretion rate in neutron star systems is an order of magnitude lower if same viscosity parameters are taken. The low transition luminosity and insignificant hysteresis in these ultracompact X-ray binaries provide further evidence that the transition luminosity is likely related to the mass in the disc. 

\end{abstract}


%%%%%%%%%%%%%%%%% BODY OF PAPER %%%%%%%%%%%%%%%%%%

\section{Introduction}
 Based on the classification in the X-ray color-color diagram, the atoll sources in neutron star (NS) low-mass X-ray binaries (LMXBs) display three distinctive spectral states,  namely the "extreme island" state,  the "island" state and the "banana" state (\citealt{Hasinger+Klis+1989,Klis+2006,Wijnands+etal+2017}). Most of those atoll sources evolve between the spectral states with time. There are many similarities in the energy spectral and timing properties of atoll sources and Galactic black hole binaries (BHBs): tentatively speaking, the "extreme island" state corresponds to  the low/hard state of BHBs when their energy spectra are both dominated  by a Comptonized (or power-law) component and their power spectra both include similar high-frequency noise with a large amplitude; the "upper banana" state roughly corresponds to the high/soft state of BHBs, which shows primarily a power-law power spectrum and a softer energy spectrum (\citealt{Klis+1994,Klis+2006,Lin+etal+2007}). To be consistent with BHBs, we will use the terms "hard" and "soft" state in this paper respectively.

To discriminate these spectral states and identify spectral state transitions from X-ray monitoring observations, it may be more straightforward to calculate the hardness ratios between the count rates in a hard X-ray band and in a soft X-ray band to  infer the relative strengths of the spectral components and identify the two major spectral states (e.g. \citealt{Yu+Dolence+2007,Yu+Yan+2009}).  As shown in \cite{Yu+Yan+2009}, the thresholds of the two spectral states indicated by the hardness ratios between the Swift/Burst Alert Telescope (15--50 keV) and the RXTE/All-Sky Monitor (2--12 keV) for black hole X-ray binaries and neutron star LMXBs can be determined from the distribution of the overall hardness ratios,  in which the two main X-ray spectral states, the hard state and the soft state, are obviously shown as two peaks due to the source concentration in the hardness ratios. The hardness ratio threshold for the hard state was taken as 1.0 for most sources and the threshold for the soft state was set to 0.2 for neutron star LMXBs, while for black hole XRBs the threshold of the soft state is a bit lower.  The weakness of using the hardness to distinguish spectral states is that, in the case of black hole binaries, the black hole intermediate state can not be clearly identified due to the degeneracy in hardness ratios. If one takes the black hole intermediate state as purely transitional stages between the two major states, then the investigation and identification of the transitions between the two major spectral states with the X-ray monitoring data can be very efficient.

The X-ray spectral state transition from the hard state to the soft state is thought to occur when the instantaneous mass accretion rate goes beyond a critical value in the popular accretion theories,  which depends on viscosity coefficient $\alpha$, pressure parameter $\beta$ and transition radius $R_{\rm tr}$ (\citealt{Esin+etal+1997}). Specifically, in the theory of Advection-Dominated Accretion Flow (ADAF), \cite{Narayan+Yi+1995} obtained that the instantaneous accretion rate of state transition is $\sim\alpha\dot{\rm M}_{\rm Edd}$ for black holes and $\sim0.1\alpha\dot{\rm M}_{\rm Edd}$ for neutron star because matter accreted onto the solid surface of neutron stars can release energy into radiation which provides additional soft photons to enhance the cooling of the advection-dominated accretion flow. However, X-ray observations in the past few decades have persistently shown that the critical accretion rate, if there is such a threshold for state transitions,  is not a constant at all. \cite{Miyamoto+etal+1995} found that the hard-to-soft transitional luminosity was larger than the luminosity corresponding to soft-to-hard transition by a factor of $\sim100$ from the observations of black hole transient GX 339-4. \cite{Homan+etal+2001} found a similar phenomenon in the black hole transient XTE J1550-564 and suggested that additional parameter should be invoked.  The same hysteresis is also seen in individual outbursts in neutron star soft X-ray transients (\citealt{Yu+etal+2003,Maccarone+Coppi+2003,Yu+Dolence+2007,Gladstone+etal+2007}).  Therefore, there should be parameters other than the mass accretion rate which determine the spectral state transitions (e.g.  \citealt{Smith+etal+2002,Zdziarski+etal+2004}).

To investigate what parameters drove the scale of the hysteresis effect, a series of study of the hard-to-soft transitions in individual black hole and neutron star soft X-ray transients (or flaring persistent X-ray sources) have been performed  (\citealt{Yu+etal+2004,Yu+Dolence+2007,Yu+etal+2007}). It has been found that the luminosity corresponding to the hard-to-soft transition correlates with the peak luminosity of the outburst or the following soft state, indicating the scale of the hysteresis is not determined by the instantaneous parameters (e.g., instantaneous mass accretion rate) but the overall properties of the outbursts, such as the mass stored in the accretion disc which is responsible for the specific outburst. Further studies of the spectral state transitions in Galactic X-ray binaries reveal the link between the rate-of-change of the mass accretion rate and the hysteresis effect, which points to non-stationary accretion regimes in BH and NS systems (\citealt{Yu+Yan+2009,Tang+etal+2011}). 

For constraining physical models, it's essential to measure the minimum mass accretion rate when the hard-to-soft spectral state transition occurs from observations, which corresponds to the critical mass accretion rate in theory.  As shown in the systematic studies of the hard-to-soft spectral state transitions detected with monitoring data from RXTE/ASM and Swift/BAT in more than 20 Galactic X-ray binaries (\citealt{Yu+Yan+2009,Tang+etal+2011}),  the lowest luminosity corresponding to the hard-to-soft spectral state transition has been seen in sources with a disc of relatively small mass, either HMXBs such as Cygnus X-1 or the Ultracompact X-ray binaries (UCXBs). UCXBs are LMXBs with orbital period less than 80 minutes. \cite{Zand+etal+2007} found the mass-transfer rate of almost all persistent UCXBs are less than $0.02~\dot{\rm M}_{\rm Edd}$ except 4U 1820-30, which possibly has the shortest orbital period of all LMXBs and shows the lowest 40--100/20--40 keV hardness ratio among all persistent UCXBs from the 2nd IBIS/ISGRI catalog (\citealt{Bird+etal+2006}).  As mentioned above, it is very important to accurately measure the X-ray luminosity corresponding to the state transitions in NS systems and compare them with those in BH systems, which will reveal the differences brought by the hard surface of neutron stars. \cite{Wu+etal+2010} found a positive correlation between orbital period and luminosity for soft X-ray transients. The study will probably shed light on the understanding of the correlation, especially the situation in the systems of short orbital periods.

%________________________________________________Observation
\section{observations}
\subsection{Samples}
To identify the hard-to-soft transitions at the lowest luminosity in LMXBs, we have surveyed all of known and candidate UCXBs and check if we can find any spectral state transitions in these sources in the all-sky X-ray monitoring data from the RXTE/ASM, the Swift/BAT and the MAXI. The initial sample includes 15 known or suspected UCXBs in the catalog of Low-Mass X-ray Binaries (LMXBs, \citealt{Liu+etal+2007}), seven additional UCXBs from \cite{Cartwright+etal+2013} and six additional candidate UCXBs identified by \cite{Zand+etal+2007}.  

In the list, we have investigated 4U 1823-00 and EXO 1745-248 first.  Their orbital periods are possibly longer than 80 minutes (\citealt{Shahbaz+etal+2007,Ferraro+etal+2015}) and their state transition luminosities (15--50 keV) are more than ${10}^{36}$ erg s$^{-1}$, so we excluded them from the following analysis. 4U 1905+000 has not been monitored by Swift/BAT probably due to its faintness , and SLX 1737-282 is neither in the monitoring catalog of RXTE/ASM nor of MAXI for the same reason.  These two sources are also lack of coverage with RXTE/PCA and Swift/XRT observations. Therefore, we excluded 4U 1905+000 and SLX 1737-282 from our samples. Swift/BAT hasn't covered any outburst from transient sources XTE J0929-314, XTE J1751-305 and XTE J1807-294, so we can't identify their state transitions based on monitoring observations. Although several outbursts from them have been captured by RXTE/PCA, \cite{Darias+etal+2014} and \cite{Gladstone+etal+2007} have analysed these observations and found that they all stay in hard state during the observations. Furthermore, there are few Swift/XRT observations of XTE J1807-294 and  XTE J0929-314 and only several Swift/XRT observations covered the 2007 outburst of XTE J1751-305. Therefore, we excluded the three transient sources as well. 

\begin{table*}
\centering
\caption{The hard-to-soft transition luminosity (15--50 keV) of 15 known or suspected UCXBs estimated from X-ray monitoring data.}
\begin{tabular}{cccccc}
\hline
Source	&	Intensity				&	Flux&	luminosity					&		Distance				&	period		\\
&(mCrab)				&	($10^{-10} ~{\rm erg~s^{-1}~cm^{-2}}$)	 &	($10^{35} ~{\rm erg~s^{-1}}$)		&	(kpc)				&	(minutes)\\
\hline	
\multicolumn{6}{c}{Persistent}\\
\hline																	
4U 1246-58	&	6.8	--	12		&	0.9	--	1.6		&	2	--	3.5		&	$	4.3	\pm	0.6^a	$	&	?		\\		
2S 0918-549	&	4.5	--	18		&	0.6	--	2.4		&	2.2	--	8.1		&	$	5.4	\pm	0.8^b	$	&	$17.4^c$	\\		
4U 0614+09	&	27	--	37		&	3.6	--	4.8		&	4.4	--	5.8		&	$	3.2	\pm	0.5^d	$	&	$51.3^e$	\\											
SAX J1712.6-3739	&	8.3	--	11		&	1.1	--	1.4		&	6.4	--	8.3		&	$	7	\pm	1.1^f	$	&	?		\\										
4U 1850-087	&	10	--	34		&	1.3	--	4.5		&	7.6	--	26		&	$	6.9	\pm	0.3^g	$	&	$20.6^h$		\\			
GRS 1724-308	&	10	--	33		&	1.3	--	4.3		&	9.2	--	29		&	$	7.5	\pm	0.4	^g$	&	?		\\\	
SLX 1744-299 (and SLX 1744-300)$^*$	&	16	--	26		&	2.1	--	3.4		&	10	--	16		&	$	6.3\pm	0.9	^i$	&	$?$		\\
4U 1543-62	&	15	--	34		&	1.9	--	4.4		&	11	--	26		&	$	7	\pm	3.5	^j$	&	$18.2^j$	\\
GX 354-0	&	30	--	137	&	3.9	--	18		&	12	--	56		&	$	5.2	\pm	0.8^k	$	&	$10.8^k$	\\
SLX 1735-269	&	11	--	19		&	1.5	--	2.5		&	12	--	21		&	$	8.5	\pm	1.3^l	$	&	?		\\	
M 15 (X-2)$^*$	&	8.6	--	17		&	1.1	--	2.3		&	14	--	29		&	$	10.4	\pm	0.5	^g$	&	$22.6^m$		\\
4U 1916-053	&	11	--	31	&	1.4	--	4.0		&	15	--	41		&	$	9.3	\pm	1.4	^n$	&	$50^o$		\\
4U 0513-40	&	8.3	--	47		&	1.1	--	6.2		&	20	--	42		&	$	12.1	\pm	0.6^g	$	&	$17^p$	\\		
H 1820-303	&	51	--	145		&	6.7	--	19		&	50	--	140		&	$	7.9	\pm	0.4	^g$	&	$11^q$		\\
\hline
\multicolumn{6}{c}{Transient}\\
\hline																		
XB 1832-330	&	11	--	27		&	1.5	--	3.5		&	16	--	36		&	$	9.3	\pm	0.5^r	$	&	$56?^s$		\\	
Swift J1756.9-2508	&	20	--	29		&	2.6	--	3.8		&	20	--	29		&	$	8	\pm	4^t	$	&	$54.7^t$		\\
RX J1709.5-2639	&	$\sim 35$			&	$\sim 4.6$			&	$\sim43	$		&	$	8.8	\pm	0.5^{g,u}	$	&	?		\\
\hline														
\end{tabular}
\begin{tablenotes}
\item{\textbf{$^*$} The flux of SLX 1744-299 or M 15 X-2 is confused by their adjacent X-ray sources. }  
\item{\textbf{Notes.}  The transition luminosity is estimated at the last hard spectral state before the transition by assuming a Crab-like energy spectrum. We take the intensity as a range when multiple hard-to-soft transition occurred during MJD 53414-57727 or as a value for single hard-to-soft transition.The errors of distance are set as 15\% for X-ray bursts and as 5\% for Globular cluster (see \citealt{Cartwright+etal+2013}). }
\item{\textbf{Ref.} $^a$\cite{Zand+etal+2008};$^b$\cite{Zand+etal+2005};$^c$\cite{Zhong+Wang+2011};$^d$\cite{Kuulkers+etal+2010};$^e$\cite{Shahbaz+etal+2008};$^f$\cite{Cocchi+etal+2001};$^g$\cite{Harris+1996};$^h$\cite{Homer+etal+1996};$^i$\cite{Zand+etal+2007};$^j$\cite{Wang+Chakrabarty+2004};$^k$\cite{Galloway+etal+2010};$^l$\cite{Molkov+etal+2005};$^m$\cite{Dieball+etal+2005};$^n$\cite{Yoshida+1993};$^o$\cite{Walter+etal+1982};$^p$\cite{Zurek+etal+2009};$^q$\cite{Stella+etal+1987};$^r$\cite{Ortolani+etal+1994};$^s$\cite{Heinke+etal+2001};$^t$\cite{Krimm+etal+2007};$^u$\cite{Jonker+etal+2003}.}
\end{tablenotes}
\label{table1}
\end{table*}

By investigating the archive data from the RXTE/ASM, the Swift/BAT and MAXI and the hardness ratios for these 21 UCXBs, we excluded the persistent sources 4U 1626-67 and 4U 1812-12 because they have stayed in hard state during the entire monitoring period of Swift/BAT.  1RXS J172525.2-325717 is excluded since it's too faint for us to identify any state transition with monitoring data and it also lacks of RXTE/PCA and Swift/XRT observations for in-depth analysis. The X-ray monitors themselves can't resolve SLX 1744-299, M 15 X-2 or NGC 6440 X-2 from their adjacent X-ray sources. The XMM-Newton observation (Obs. ID: 0152920101) has resolved SLX 1744-299 from another X-ray burster SLX 1744-300 and found that the flux of SLX 1744-299 is about twice higher than SLX 1744-300 in 0.5--10 keV energy band (\citealt{Mori+etal+2005}). Similarly, M 15 X-2 is responsible for most X-ray emission of the flare in 2011 (\citealt{Sivakoff+etal+2011}) and M 15 X-2 is typically more bright than its adjacent LMXB, AC 211 (\citealt{Hannikainen+etal+2005}).  Therefore, SLX 1744-299 and M 15 X-2 possibly still dominates the state transition rather than their adjacent X-ray sources. On the other hand,  the major outbursts from NGC 6440 were contributed by accreting millisecond X-ray pulsar, SAX J1748.9-2021 (\citealt{Altamirano+etal+2008,Patruno+etal+2010}). The estimated transition flux for NGC 6440 X-2 by monitoring observations is actually contributed by SAX J1748.9-2021. Therefore, we included the persistent source SLX 1744-299 and M 15 X-2 in our samples but excluded the transient source NGC 6440 X-2.

\subsection{All-sky monitoring observations}
We took the monitoring data of Swift/BAT, RXTE/ASM and MAXI from public archives (\citealt{Krimm+etal+2013}).  We obtained the orbital light curves of Swift/BAT at 15--50 keV in the period from MJD 53414 to MJD 57727, RXTE/ASM at 2--12 keV in the period from MJD 53414 (when Swift/BAT monitoring started) to MJD 55924 and those of MAXI in 2--4 keV and 4--10 keV in the period from MJD 55058 to MJD 57727.  All the raw data is in units of counts s$^{-1}$ cm$^{-2}$. We binned all orbital data into daily averaged data, and then converted them in units of Crab in order to make comparison among the different instruments. The MAXI data in 2--4 keV and 4--10 keV were combined together to match the RXTE/ASM data.  In our investigations,  we have ignored the data points whose significance is less than 1$\sigma$. The hardness ratios were calculated by the intensity in the Swift/BAT band (15--50 keV) divided by intensity in the RXTE/ASM under 12 keV or MAXI under 10 keV.  Note that, RXTE/ASM and MAXI were both operating from MJD 55058 to MJD 55924. In that period, we tended to use the data from MAXI rather than RXTE/ASM since the sensitivity of MAXI was usually better.

We identified state transitions in all our samples in the period from MJD 53414 to MJD 57727 by comparing the hardness ratios to the threshold (\citealt{Yu+Yan+2009}). Since all samples are neutron stars LMXBS, we used 0.2 as the threshold for the soft state and 1.0 for the hard state as in the previous approach (\citealt{Yu+Yan+2009}). The last data point before a source leaving the hard state is marked as the start of hard-to-soft transition.  This is because it is relatively easy to estimate the X-ray luminosity in the hard state because the energy spectrum is characterised by a  power-law component. We estimated the source X-ray flux by assuming that hard state X-ray spectra of theses sources are roughly similar to that of the Crab Nebula. Typically, the spectrum of the Crab Nebula in 2--10 keV and in 10--50 keV can be described by an absorbed power-law with index of 2.07 and 2.12 respectively  and $N_{\rm H}=4.5\times10^{21}~{\rm cm^{-2}}$ (\citealt{Kirsch+etal+2005}). Using webPIMMs v4.8c, we found that the unabsorbed flux varies by at most 20\% even if the power-law index varies by 1. However,  if $N_{\rm H}$ is up to 10 times larger, the unabsorbed flux at 2--10 or 2--12 keV energy will differs over 30\% while the unabsorbed flux at 15--50 keV energy is almost unaffected.  Therefore, we chose 15--50 keV energy band to compare the transition luminosities between those UCXBs with BAT data. Combined with the result of \cite{Kirsch+etal+2005} , we estimated the flux of the Crab nebula as $2.19\times10^{-8}$ erg s$^{-1}$ cm$^{-2}$ in 2--10 keV, $2.42\times10^{-8}$ erg s$^{-1}$ cm$^{-2}$ in 2--12 keV, and $1.31\times10^{-8}$ erg s$^{-1}$ cm$^{-2}$ in 15--50 keV, respectively.  

The estimates of the luminosity (15--50 keV) of the hard-to-soft state transitions are listed in the table~\ref{table1}. In the end, we have found hard-to-soft spectral state transitions with the lowest luminosity occurred in the three sources, namely 4U 0614+091, 2S 0914-549 and 4U 1246-588. We then focused on the spectral transitions in these sources and made use of pointed X-ray observations to derive the transition luminosity. 

\begin{figure}
	\centering
	\includegraphics[width=7cm]{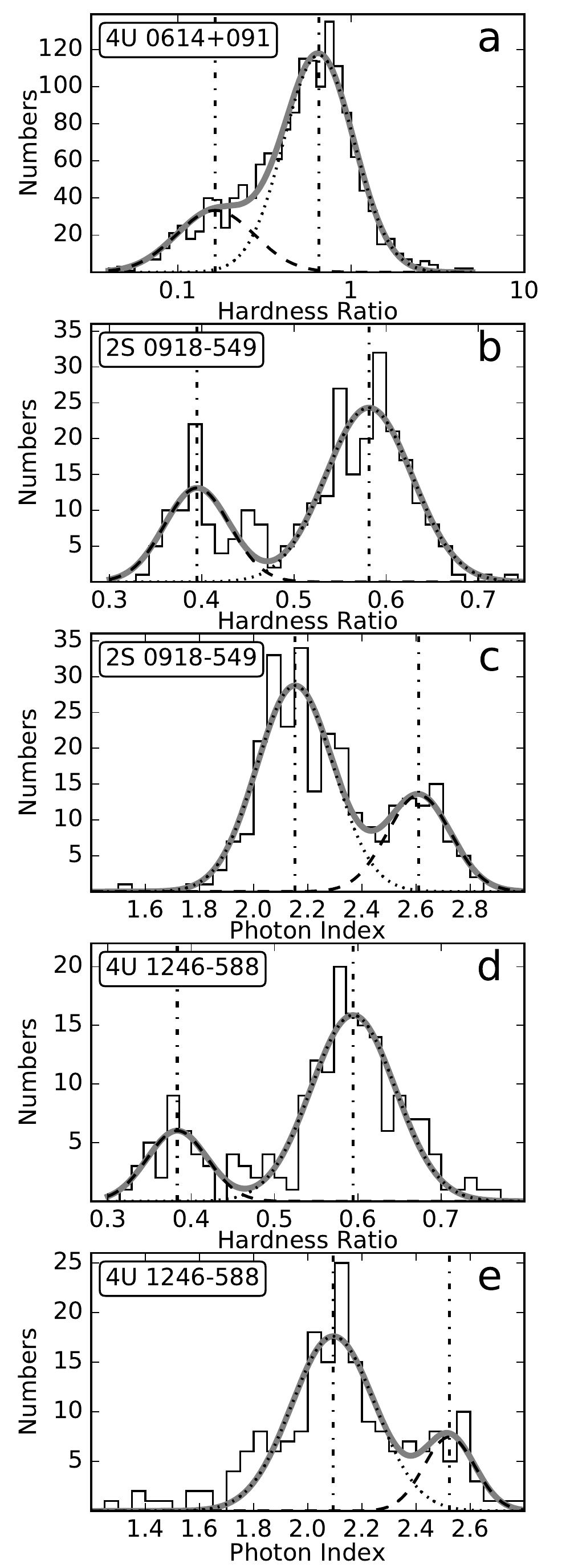}
    \caption{The distributions of the hardness ratio and the best-fitting power-law indices for the three UCXBs. The thick solid line represents the the best-fitting double Gaussian model while the dotted and the dashed line represent the individual Gaussian components.  The dashed-dotted lines indicate the centre of Gaussian components. (a) The distribution of the BAT (15--50 keV)/MAXI (2--10 keV) hardness ratio for 4U 0614+091 on logarithmic scale. The total number of hardness ratios is 1538. (b) The distribution of the 10--16 keV/6--10 keV hardness ratios from 281 RXTE/PCA observations for 2S 0918-549. (c) The distribution of best-fitting power-law indices from 281 RXTE/PCA observations for 2S 0918-549. (d) The distribution of the 10--16 keV/6--10 keV hardness ratios from 186 RXTE/PCA observations for 4U 1246-588.  (e) The distribution of best-fitting power-law indices from 186 RXTE/PCA observations for 4U 1246-588. } 
    \label{fig1}
\end{figure}

\subsection{Identification of spectral state transitions and the corresponding observations}
We measured the exact X-ray fluxes from 4U 0614+091, 2S 0914-549 and 4U 1246-588 by RXTE or Swift pointed observations, since these three sources showed the lowest hard-to-soft transition luminosity in table~\ref{table1} and the their state transitions were already covered by pointed observations of the Swift/XRT and/or the RXTE/PCA.

4U 0614+091 is the closest UCXB as its distance is 3.2 kpc (\citealt{Brandt+etal+1992}). Its accretion disc could be rich of C/O (\citealt{Nelemans+etal+2006}) and the broad emission feature at 0.6--0.7 keV has been reported (\citealt{Piraino+etal+1999};\citealt{Juett+etal+2001}). The donor of 4U 0614+091 could be a C/O white dwarf since He and H lines are lacking in its optical spectra (\citealt{Werner+etal+2006}). For 4U 0614+091, the spectral states are determined by hardness ratios between the Swift/BAT and MAXI data. Notice that, the hardness ratio 0.2 and 1.0 are not  accurate enough for the identification of spectral states in individual sources. Therefore, the spectral state thresholds for 4U 0614+091 are reset by fitting the distribution of the hardness ratio. The distribution of the hardness ratio shows a main peak and a remarkable tail over lower hardness range in the logarithmic scale (see Figure~\ref{fig1}), for the monitoring data of BAT and MAXI in the period between MJD 55058 and MJD 57727. We used a model composed of two Gaussian functions to fit the distribution.  The best-fitting parameters suggested the thresholds in the hardness ratios for the hard and the soft state respectively.  As the result, the hard state hardness ratio threshold was taken as 0.65 and the soft state hardness ratio threshold was taken as 0.16.

\begin{figure*}
  \centering
	\includegraphics[width=14cm]{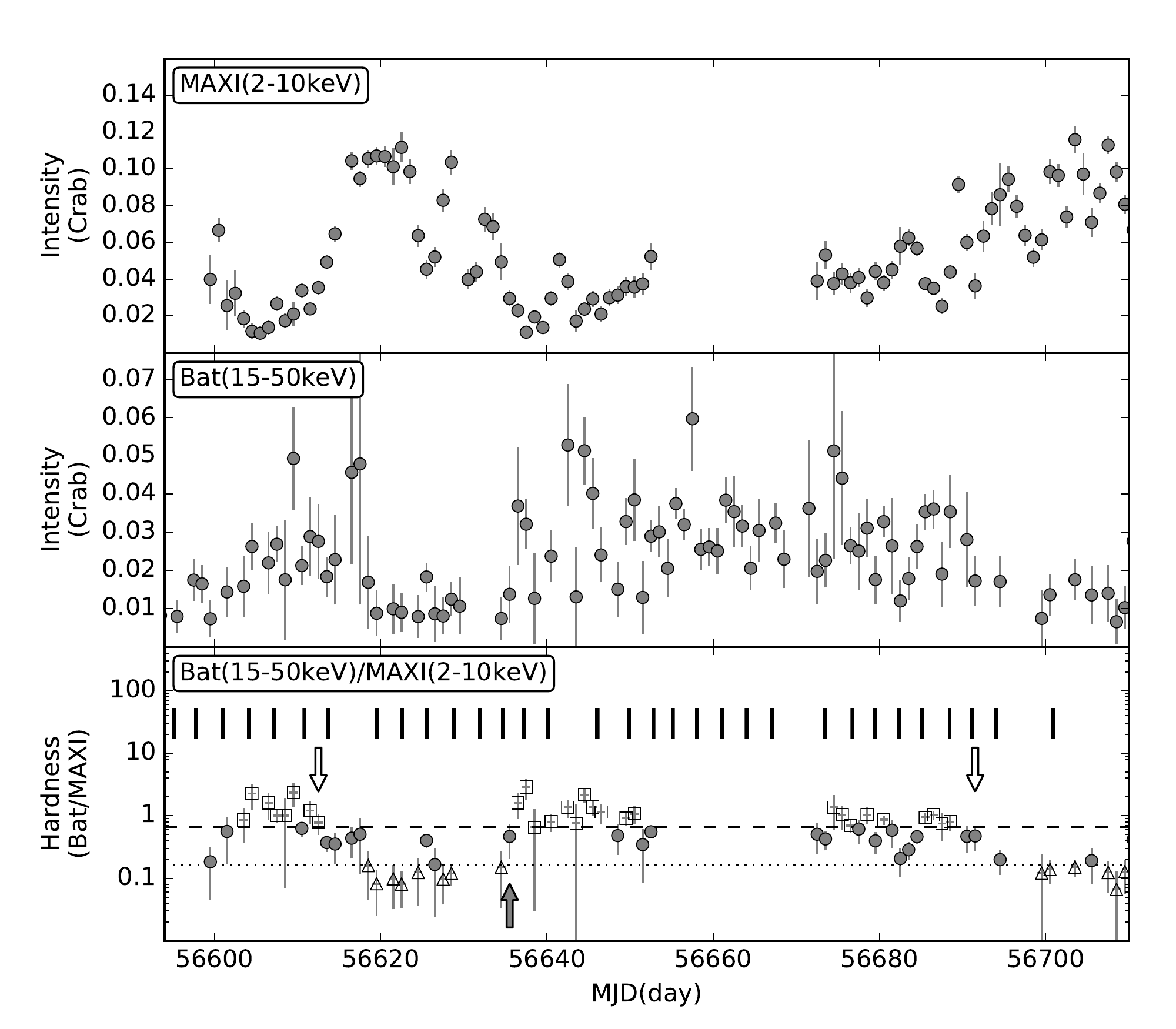}
    \caption{The monitoring light curve of 4U 0614+091 in the period from MJD 56595 to MJD 56710.  In the bottom panel, the dashed line and the dotted line represent the thresholds corresponding to the soft and the hard state, respectively. The empty arrows indicate the start of hard-to-soft transition and the filled arrows point the end of soft-to-hard transition. The vertical bars mark the time of Swift/XRT observations.
    }
    \label{fig2}
\end{figure*}

We identified state transitions between MJD 56595 and MJD 56710 during which there were rather intense monitoring observations with the Swift/XRT which allowed us to measure the transition flux. We determined the hard-to-soft transition flux by measuring the flux of last hard state observation just before the hardness ratio departed from the threshold definitely. From the Figure~\ref{fig2}, there were two XRT target-of-opportunity (TOO) observations, 00030812012 and 00030812036, triggered almost exactly at the start of hard-to-soft transition. The hard-to-soft transition flux of 4U 0614+091 could be measured from these two observations.  Similarly,  we determined the soft-to-hard transition flux by measuring the flux corresponding to first hard state observation just after the hardness ratio evolved back to the threshold of the hard state definition. We found the observation 00030812018 was taken exactly at the end of soft-to-hard transition and then measured the soft-to-hard transition flux from this observation. Note that the measured flux should be lower than the transition flux, in the assumption that the hard-to-soft transition occur during a luminosity rise  and the soft-to-hard transition occur during a luminosity decrease. 

\begin{figure*}
  \centering
	\includegraphics[width=14cm]{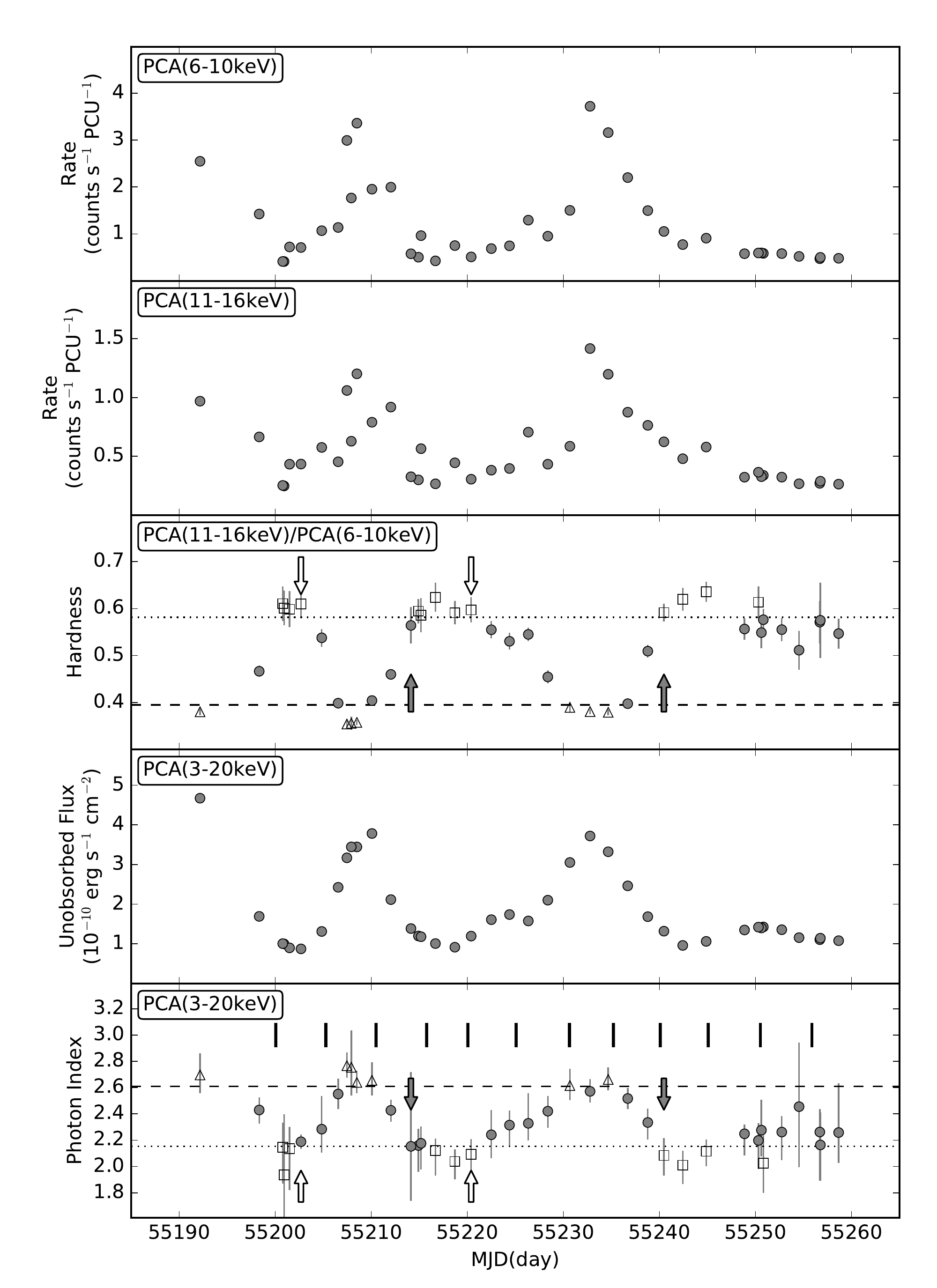}
    \caption{The X-ray monitoring observations and pointed RXTE/PCA observations of 2S 0914-549 in the period from MJD 55190 to MJD 55260. The X-ray fluxes and photon indices are measured with the observations of RXTE/PCA in the energy range 3--20 keV. The dashed line and the dotted line represent the thresholds corresponding to the soft and the hard state, respectively. The empty arrows indicate the start of the hard-to-soft transition and the filled arrows point at the end of soft-to-hard transition.The vertical bars mark the time of individual Swift/XRT observations.} 
    \label{fig3}
\end{figure*}

\begin{figure*}
  \centering
	\includegraphics[width=14cm]{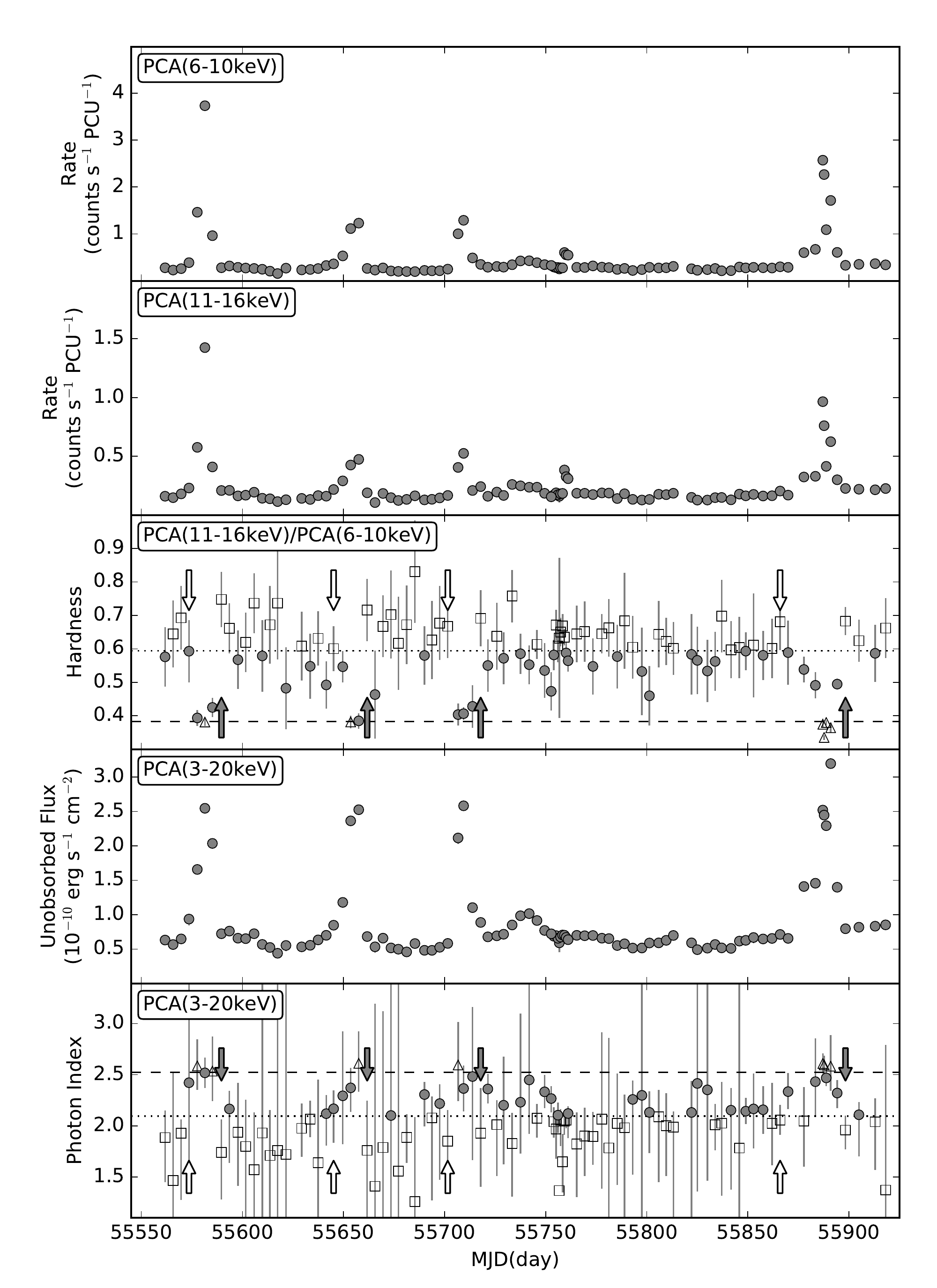}
    \caption{The X-ray monitoring observations and pointed RXTE/PCA observations of 4U 1246-588 in the period from MJD 55550 to MJD 55920. As in the previous plot, the X-ray fluxes and photon indices are measured with the observations of RXTE/PCA in the energy range 3--20 keV.}
    \label{fig4}
\end{figure*}

2S 0918-549 is similar to 4U 0614+091, which harbors a C/O white dwarf and a He-deficient accretion disc (\citealt{Deloye+Bildsten+2003}; \citealt{Nelemans+etal+2004}; \citealt{Juett+etal+2001}). 4U 1246-588 was identified as UCXB due to the lack of hydrogen spectral features from an observation with the  Very Large Telescope (\citealt{Zand+etal+2008}). The average X-ray intensity of 2S 0914-549 and 4U 1246-588 is usually lower than that of 4U 0614+091 by about one order of magnitude. This means that it's more difficult to identify the state transitions in the X-ray monitoring data. Therefore, their spectral states are further relied on the investigation of densely monitoring observations performed with the RXTE/PCA, usually performed every 2--3 days. 

We used the RXTE/PCA Standard-2 mode (STD2) data for hardness calculation and spectral analysis. The hardness ratio was defined as the ratio of count rates between 10--16 keV and 6--10 keV because the energy band minimizes the effect of interstellar absorption and avoids the complex soft spectral region (see \citealt{Darias+etal+2014} ).  Similar to the analysis for 4U 0614+091, we fitted the distributions of these hardness ratios to determine the thresholds of spectral states for these two sources. The distributions and best-fitting models are shown in the Figure~\ref{fig1}. The hardness ratio thresholds corresponding to hard state and soft state are 0.58 and 0.39 for 2S 0918-549 and are 0.59 and 0.38 for 4U 1246-588 respectively. We adopted standard RXTE/PCA spectra generated from STD2 and applied standard response and background files in the spectral analysis. We fitted all the standard PCA spectra of 2S 0914-549 and 4U 1246-588 in the energy range 3--20 keV with a model composed of a power-law component plus a simple blackbody with a fixed hydrogen column density ($N_{\rm H}$). The $N_{\rm H}$ was set to $3.0\times10^{21}$ cm$^{-2}$ for 2S 0914-549 (\citealt{Juett+Chakrabarty+2003}) and $2.5\times10^{21}$ cm$^{-2}$ for 4U 1246-588 (\citealt{Zand+etal+2008}).  the spectral state thresholds for 2S 0914-549 and 4U 1246-588 were determined by fitting the distribution of  the best-fitting power-law index (PLI) with a model composed of two Gaussians (Figure~\ref{fig1}). Based on those spectral fits, the hard state PLI thresholds were taken as 2.15 for 2S 0914-549 and as 2.09 for 4U 1246-588, respectively, and the soft state PLI thresholds were taken as 2.61 for 2S 0914-549 and as 2.52 for 4U 1246-588.  

In this way, we selected the state transitions which occurred in the period from MJD 55190 to MJD 55260 for 2S 0914-549 and from MJD 55550 to MJD 55920 for 4U 1246-588, when there were intense coverage of pointed RXTE observations with the interval as short as 2--3 days and 4--5 days, respectively, which are short enough as compared with the time-scale of the spectral state transitions. During the period between MJD 55203 and MJD 55208 and the period between MJD 55220 and MJD 55232 (Figure~\ref{fig3}), 2S 0918-549 left from the hard state since its hardness ratio decreased and spectral index went up with time obviously. 
Two hard-to-soft transitions occurred at around MJD 55203 and MJD 55220, respectively.  Similarly, two soft-to-hard transitions ended at around MJD 55214 and MJD 55240, respectively. There were a Swift/XRT observation (ID: 00031569005) performed at around MJD 55220 and another Swift/XRT observation (ID: 00031569010) performed at around MJD 55240, allowing us to perform a precise measurement of the transition flux by fitting the joint X-ray spectra from the Swift/XRT and the RXTE/PCA. 

As shown in the Figure~\ref{fig4}, for 4U 1246-588, we determined the hard-to-soft transitions at around MJD 55574, MJD 55645, MJD 55701 and MJD 55865 and the soft-to-hard transitions at around MJD 55590, MJD 55662, MJD 55718 and MJD 55898, respectively. We then used the corresponding RXTE/PCA observations to derive the transition fluxes. 

%________________________________________________Result
\section{Data analysis and results}
As introduced above, we then identified the corresponding Swift/XRT observations for 4U 0614+091 and 2S 0918-549 with the observation IDs 00030812012, 00030812018, 00030812036, 00031569005 and  00031569010 respectively. All of these observations were obtained in window timing (WT) mode which avoided significant pile-up effect. We have analysed the observations in the standard way. All the raw data were reduced by running the \emph{xrtpipeline} task of \emph{HEASOFT} (version 6.19). We used \emph{Xselect} (version 2.4d) to extract source and background spectra from the reduced data by applying a region of a circle with a radius of 30 arcsec (for source) and a concentric annulus with an inner radius of 60 arcsec and an outer radius of 120 arcsec (for background).  The extracted energy spectra were grouped into 20 counts per spectral bin. We adopted the corresponding response matrix files (RMF) from the Swift/XRT calibration data base (CALDB, version 20150721) and created ancillary response files (ARF) from exposure maps using \emph{xrtmkarf} task. 

We fitted the spectra of RXTE/PCA and Swift/XRT observations using \emph{XSPEC 12.9.0n} and obtained the best-fitting parameters and the X-ray fluxes. We randomized the spectral parameters using the best-fitting values and errors by 10000 times and recorded the resulting ratios of the bolometric flux (in 0.1--100 keV) to the measured flux. The bolometric flux is estimated from spectral parameters based on the assumption that the power-law spectral component is up to around 100 keV and down to around 0.1 keV. We fitted the ratio distributions with Gaussians with different widths below and above the centre of the Gaussians and the uncertainties of bolometric correlation factors were derived from the best-fitting "Gaussian" models following \cite{Kuulkers+etal+2010}.

We inferred the luminosities from the fluxes by assuming the distances as $3.2\pm0.5$ kpc  for 4U 0614+091,  $5.4\pm0.8$ kpc for 2S 0918-549 and $4.3\pm0.6$ kpc for 4U 1246-588 and estimated the bolometric luminosities in Eddington unit by the bolometric correction factors. The Eddington luminosity is calculated by assuming that the mass of neutron star  is $1.4~{\rm M}_{\odot}$. Therefore, the Eddington luminosity is $2.0\times10^{38}{\rm ~erg~s^{-1}}$ for a hydrogen-rich photosphere and $3.5\times10^{38}{\rm ~erg~s^{-1}}$ for a hydrogen-poor photosphere.  The estimated uncertainty in the Eddington luminosity is about 30\%(\citealt{Zand+etal+2007}). Finally, the mass accretion rates are derived from the bolometric luminosities by assuming the radiative efficiency as 0.1 for NS LMXBs (\citealt{Narayan+McClintock+2008}). All results are showed in table~\ref{table2}, and the uncertainties represent 90\% confidence limits.

\subsection{4U 0614+091}
We fitted the XRT spectra of 4U 0614+091 {in 0.6--10 keV energy} with a model composed of a power-law component, accounting for the Comptonized X-ray emission, plus a blackbody component representing the X-ray emission from the surface of the neutron star or boundary layer, with an additional Gaussian component around 0.6 keV, which represents an oxygen line that was identified before (\citealt{Piraino+etal+1999}). We fixed the centre of Gaussian emission line at 0.62 keV following previous studies (\citealt{Juett+etal+2001,Migliari+etal+2010}). The best-fitting parameters are listed in the table~\ref{table2}.

The unabsorbed fluxes of 4U 0614+091 in 0.6--10 keV are $(2.8\pm0.2)\times10^{-9}$ erg s$^{-1}$ cm$^{-2}$ and $(3.1\pm0.1)\times10^{-9}$ erg s$^{-1}$ cm$^{-2}$ , respectively, from the two observations corresponding to the hard-to-soft transition at around MJD 56613.7 and MJD 56691.1. Therefore, the lower limit of the luminosity corresponding to the hard-to-soft transition in 4U 0614+091 is $(3.4 \pm 1.1) \times10^{36}$ erg s$^{-1}$ taking the distance as $3.2\pm0.5$ kpc. Based on the best-fitting model, we estimated the bolometric correction factor as $2.5^{+0.3}_{-0.1}$. So the lowest bolometric luminosity (in 0.1--100 keV) of the hard-to-soft transition is $(8.6\pm 2.8)\times10^{36}$ erg s$^{-1}$, corresponding to a mass accretion rate of $0.043\pm0.019~\dot{\rm M}_{\rm Edd}$ for a hydrogen-rich photosphere.
Similarly, the soft-to-hard transition flux at around MJD 56634.7 in 0.6--10 keV is $(3.0^{+0.4}_{-0.2})\times10^{-9}$ erg s$^{-1}$ cm$^{-2}$ and the corresponding mass accretion rate is $0.050^{+0.023}_{-0.022}~\dot{\rm M}_{\rm Edd}$ for a hydrogen-rich photosphere. The soft-to-hard transition luminosity is almost consistent with the hard-to-soft transition luminosity. Note that, for a hydrogen-poor photosphere which is generally true for UCXBs, the actual  accretion rate in Eddington unit should be lower by half.  This yield the mass accretion rates corresponding to hard-to-soft and soft-to-hard transition are $0.024\pm0.011~\dot{\rm M}_{\rm Edd}$ and $0.029\pm0.013~\dot{\rm M}_{\rm Edd}$.

\begin{sidewaystable*}
\caption{The best-fitting parameters and the inferred luminosities from the observations of Swift/XRT and RXTE/PCA. All errors are quoted at 90\% confidence levels. }
\scriptsize
\begin{tabular}{cccccccccccc}
\hline\hline
\multicolumn{12}{c}{Hard-to-soft transition}\\
\hline
Parameters&&\multicolumn{2}{c}{4U 0614+091}&&\multicolumn{2}{c}{2S 0918-549}&&\multicolumn{4}{c}{4U 1246-588}\\
\cline{1-1}\cline{3-4} \cline{6-7} \cline{9-12}
Obs. ID	&	&	00030812012	&	00030812036	&	&	95343-01-03-00 	&	00031569005+95343-01-12-00 	&	&	96340-01-04-00 	&	96340-01-22-00 	&	96340-01-36-00 	&	96340-01-77-00 	\\
Device	&	&	Swift/XRT	&	Swift/XRT	&	&	RXTE/PCA	&	Swift/XRT+RXTE/PCA	&	&	RXTE/PCA	&	RXTE/PCA	&	RXTE/PCA	&	RXTE/PCA	\\
Time (MJD)	&	&	56613.7	&	56691.08	&	&	55202.68	&	55220.4	&	&	55573.59	&	55645.12	&	55701.57	&	55865.94	\\
Model	&	&	wabs*(pl+bb+ga)	&	wabs*(pl+bb+ga)	&	&	wabs*pl	&	wabs*(pl+bb+ga)	&	&	wabs*pl	&	wabs*pl	&	wabs*pl	&	wabs*pl	\\
Energy band	&	&	0.6--10 keV	&	0.6--10 keV	&	&	3--20 keV	&	0.6--20 keV	&	&	3--20 keV	&	3--20 keV	&	3--20 keV	&	3--20 keV	\\
$\chi^2$/dof	&	&	534.47/537  	&	646.11/635  	&	&	30.31/38  	&	262.71/265  	&	&	15.01/38  	&	13.76/38  	&	30.82/38  	&	25.71/38  	\\
$N_{\rm H}$ ($\rm \times10^{22}~cm^{-2}$)	&	&	 $0.383^{+0.036}_{-0.051}$ 	&	 $0.404\pm0.027$ 	&	&	 $0.3$(fixed) 	&	 $0.284^{+0.05}_{-0.051}$ 	&	&	 $0.25$(fixed) 	&	 $0.25$(fixed) 	&	 $0.25$(fixed) 	&	 $0.25$(fixed) 	\\
$\Gamma$	&	&	 $2.36\pm0.12$ 	&	 $2.16^{+0.05}_{-0.06}$ 	&	&	 $2.19^{+0.06}_{-0.05}$ 	&	 $2.07\pm0.07$ 	&	&	 $2.47\pm0.17$ 	&	 $2.25^{+0.13}_{-0.12}$ 	&	 $2.07^{+0.18}_{-0.17}$ 	&	 $2.05^{+0.15}_{-0.14}$ 	\\
$N_{\rm pl}(\times10^{-2})$	&	&	 $69.1^{+5.3}_{-9.1}$ 	&	 $62.9^{+5.9}_{-5.5}$ 	&	&	 $4.2^{+0.43}_{-0.39}$ 	&	 $4.41^{+0.58}_{-0.59}$ 	&	&	 $7.84^{+2.65}_{-1.95}$ 	&	 $4.49^{+1.1}_{-0.88}$ 	&	 $2.13^{+0.79}_{-0.57}$ 	&	 $2.61^{+0.79}_{-0.6}$ 	\\
kT(keV)	&	&	 $1.08^{+0.36}_{-0.61}$ 	&	 $0.386^{+0.023}_{-0.025}$ 	&	&	-	&	 $0.442^{+0.169}_{-0.442}$ 	&	&	-	&	-	&	-	&	-	\\
$N_{\rm bb}(\times10^{-4})$	&	&	 $9.69^{+12.6}_{-8.65}$ 	&	 $31.5^{+7.3}_{-7.6}$ 	&	&	-	&	 $1.05^{+1.04}_{-1.05}$ 	&	&	-	&	-	&	-	&	-	\\
$E_{\rm ga}$(keV)	&	&	 $0.62$(fixed) 	&	 $0.62$(fixed) 	&	&	-	&	 $0.635$(fixed) 	&	&	-	&	-	&	-	&	-	\\
$\sigma_{\rm ga}$(keV)	&	&	 $0.112^{+0.007}_{-0.006}$ 	&	 $0.129\pm0.005$ 	&	&	-	&	 $0.108^{+0.061}_{-0.108}$ 	&	&	-	&	-	&	-	&	-	\\
$N_{\rm ga}(\times10^{-2})$	&	&	 $57.3^{+17.9}_{-18.4}$ 	&	 $64.9^{+13.3}_{-11.6}$ 	&	&	-	&	 $0.88^{+1.193}_{-0.811}$ 	&	&	-	&	-	&	-	&	-	\\
Unabsorbed flux($\times10^{-10}$ erg s$^{-1}$ cm$^{-2}$) &	&	 $28.0^{+2.0}_{-2.4}$ 	&	 $31.3^{+1.5}_{-1.4}$ 	&	&	 $0.873\pm0.022$ 	&	 $2.41^{+0.15}_{-0.12}$ 	&	&	 $0.934^{+0.068}_{-0.067}$ 	&	 $0.831\pm0.047$ 	&	 $0.559^{+0.046}_{-0.045}$ 	&	 $0.712^{+0.049}_{-0.048}$ 	\\
Luminosity($\times10^{35}$ erg/s)$^a$	&	&	 $34.2^{+11.0}_{-11.1}$ 	&	 $38.2\pm12.1$ 	&	&	 $3.04\pm0.9$ 	&	 $8.38^{+2.53}_{-2.52}$ 	&	&	 $2.06\pm0.59$ 	&	 $1.83\pm0.52$ 	&	 $1.23\pm0.36$ 	&	 $1.57\pm0.45$ 	\\
Bolometric correction factor$^b$	&	&	 $2.5^{+0.29}_{-0.14}$ 	&	 $2.25\pm0.04$ 	&	&	 $4.6^{+0.47}_{-0.35}$ 	&	 $1.96^{+0.07}_{-0.04}$ 	&	&	 $7.14^{+4.66}_{-1.77}$ 	&	 $4.81^{+1.51}_{-0.71}$ 	&	 $3.63^{+1.13}_{-0.4}$ 	&	 $3.64^{+0.88}_{-0.32}$ 	\\
$L_{\rm bol}$($\times10^{-3}~{\rm L}_{\rm Edd,H-rich}$)$^{a,c}$	&	&	 $42.7^{+19.4}_{-19.0}$ 	&	 $42.9\pm18.7$ 	&	&	 $6.98^{+3.04}_{-3.0}$ 	&	 $8.2^{+3.51}_{-3.49}$ 	&	&	 $7.35^{+5.69}_{-3.56}$ 	&	 $4.4^{+2.28}_{-1.93}$ 	&	 $2.24^{+1.17}_{-0.97}$ 	&	 $2.86^{+1.37}_{-1.21}$ 	\\
$L_{\rm bol}$($\times10^{-3}~{\rm L}_{\rm Edd,H-poor}$)$^{a,c}$	&	&	 $24.4^{+11.1}_{-10.9}$	&	 $24.5\pm10.7$	&	&	 $3.99^{+1.73}_{-1.71}$	&	 $4.69^{+2.0}_{-1.99}$	&	&	 $4.2^{+3.25}_{-2.03}$	&	 $2.52^{+1.31}_{-1.1}$	&	 $1.28^{+0.67}_{-0.55}$	&	 $1.63^{+0.78}_{-0.69}$	\\
\hline\hline
\multicolumn{12}{c}{Soft-to-hard transition}\\
\hline
Parameters&&\multicolumn{2}{c}{4U 0614+091}&&\multicolumn{2}{c}{2S 0918-549}&&\multicolumn{4}{c}{4U 1246-588}\\
\cline{1-1}\cline{3-4} \cline{6-7} \cline{9-12}
Obs. ID	&	&	\multicolumn{2}{c}{	00030812018	}	&	&	95343-01-09-00 	&	00031569010+95343-01-22-00 	&	&	96340-01-08-00 	&	96340-01-26-00 	&	96340-01-40-00 	&	96340-01-79-00 	\\
Device	&	&	\multicolumn{2}{c}{	Swift/XRT	}	&	&	RXTE/PCA	&	Swift/XRT+RXTE/PCA	&	&	RXTE/PCA	&	RXTE/PCA	&	RXTE/PCA	&	RXTE/PCA	\\
Time (MJD)	&	&	\multicolumn{2}{c}{	56634.71	}	&	&	55214.91	&	55240.47	&	&	55589.63	&	55661.72	&	55717.86	&	55898.27	\\
Model	&	&	\multicolumn{2}{c}{	wabs*(pl+bb+ga)	}	&	&	wabs*pl	&	wabs*(pl+bb+ga)	&	&	wabs*pl	&	wabs*pl	&	wabs*pl	&	wabs*pl	\\
Energy band	&	&	\multicolumn{2}{c}{	0.6--10 keV	}	&	&	3--20 keV	&	0.6--20 keV	&	&	3--20 keV	&	3--20 keV	&	3--20 keV	&	3--20 keV	\\
$\chi^2$/dof	&	&	\multicolumn{2}{c}{	743.99/616  	}	&	&	19.03/38  	&	244.22/263  	&	&	26.15/38  	&	23.09/38  	&	20.47/38  	&	36.95/38  	\\
$N_{\rm H}$ ($\rm \times10^{22}~cm^{-2}$)	&	&	\multicolumn{2}{c}{	 $0.488^{+0.028}_{-0.026}$ 	}	&	&	 $0.3$(fixed) 	&	 $0.279^{+0.053}_{-0.049}$ 	&	&	 $0.25$(fixed) 	&	 $0.25$(fixed) 	&	 $0.25$(fixed) 	&	 $0.25$(fixed) 	\\
$\Gamma$	&	&	\multicolumn{2}{c}{	 $2.48^{+0.09}_{-0.08}$ 	}	&	&	 $2.26\pm0.05$ 	&	 $2.1\pm0.06$ 	&	&	 $2.03^{+0.14}_{-0.13}$ 	&	 $2.15^{+0.16}_{-0.15}$ 	&	 $2.17\pm0.14$ 	&	 $2.13\pm0.08$ 	\\
$N_{\rm pl}(\times10^{-2})$	&	&	\multicolumn{2}{c}{	 $67.1^{+3.8}_{-3.4}$ 	}	&	&	 $6.59^{+0.59}_{-0.54}$ 	&	 $5.02^{+0.69}_{-0.64}$ 	&	&	 $2.44^{+0.65}_{-0.52}$ 	&	 $2.91^{+0.9}_{-0.68}$ 	&	 $3.98^{+1.12}_{-0.87}$ 	&	 $3.31^{+0.48}_{-0.42}$ 	\\
kT(keV)	&	&	\multicolumn{2}{c}{	 $1.32\pm0.1$ 	}	&	&	-	&	 $0.729^{+0.158}_{-0.091}$ 	&	&	-	&	-	&	-	&	-	\\
$N_{\rm bb}(\times10^{-4})$	&	&	\multicolumn{2}{c}{	 $23.5^{+7.8}_{-7.9}$ 	}	&	&	-	&	 $1.96^{+1.12}_{-1.07}$ 	&	&	-	&	-	&	-	&	-	\\
$E_{\rm ga}$(keV)	&	&	\multicolumn{2}{c}{	 $0.62$(fixed) 	}	&	&	-	&	 $0.635$(fixed) 	&	&	-	&	-	&	-	&	-	\\
$\sigma_{\rm ga}$(keV)	&	&	\multicolumn{2}{c}{	 $0.117\pm0.004$ 	}	&	&	-	&	 $0.109^{+0.029}_{-0.028}$ 	&	&	-	&	-	&	-	&	-	\\
$N_{\rm ga}(\times10^{-2})$	&	&	\multicolumn{2}{c}{	 $75.0^{+15.7}_{-13.0}$ 	}	&	&	-	&	 $2.44^{+1.75}_{-1.22}$ 	&	&	-	&	-	&	-	&	-	\\
Unabsorbed flux($\times10^{-10}$ erg s$^{-1}$ cm$^{-2}$) &	&	\multicolumn{2}{c}{	 $30.3^{+3.9}_{-2.3}$ 	}	&	&	 $1.18\pm0.03$ 	&	 $2.83^{+0.21}_{-0.16}$ 	&	&	 $0.695^{+0.044}_{-0.043}$ 	&	 $0.656^{+0.046}_{-0.045}$ 	&	 $0.863^{+0.056}_{-0.055}$ 	&	 $0.769\pm0.027$ 	\\
Luminosity($\times10^{35}$ erg/s)$^a$	&	&	\multicolumn{2}{c}{	 $37.0^{+12.5}_{-11.9}$ 	}	&	&	 $4.12\pm1.22$ 	&	 $9.83^{+3.0}_{-2.97}$ 	&	&	 $1.53\pm0.44$ 	&	 $1.45\pm0.42$ 	&	 $1.9\pm0.54$ 	&	 $1.7\pm0.48$ 	\\
Bolometric correction factor$^b$	&	&	\multicolumn{2}{c}{	 $2.69^{+0.24}_{-0.17}$ 	}	&	&	 $5.23^{+0.53}_{-0.42}$ 	&	 $1.93^{+0.07}_{-0.04}$ 	&	&	 $3.59^{+0.72}_{-0.25}$ 	&	 $4.06^{+1.28}_{-0.52}$ 	&	 $4.21^{+1.28}_{-0.54}$ 	&	 $4.2^{+0.55}_{-0.36}$ 	\\
$L_{\rm bol}$($\times10^{-3}~{\rm L}_{\rm Edd,H-rich}$)$^{a,c}$	&	&	\multicolumn{2}{c}{	 $49.9^{+23.0}_{-22.2}$ 	}	&	&	 $10.8^{+4.7}_{-4.6}$ 	&	 $9.49^{+4.07}_{-4.04}$ 	&	&	 $2.75^{+1.27}_{-1.16}$ 	&	 $2.93^{+1.53}_{-1.28}$ 	&	 $4.0^{+2.06}_{-1.74}$ 	&	 $3.56^{+1.54}_{-1.5}$ 	\\
$L_{\rm bol}$($\times10^{-3}~{\rm L}_{\rm Edd,H-poor}$)$^{a,c}$	&	&	\multicolumn{2}{c}{	 $28.5^{+13.1}_{-12.7}$	}	&	&	 $6.16^{+2.67}_{-2.64}$	&	 $5.42^{+2.33}_{-2.31}$	&	&	 $1.57^{+0.72}_{-0.66}$	&	 $1.68^{+0.87}_{-0.73}$	&	 $2.29^{+1.18}_{-0.99}$	&	 $2.04^{+0.88}_{-0.86}$	\\
\hline\hline
\end{tabular}

\begin{tablenotes}
\item{

{$^a$ The estimated luminosities are based on the assumption that the distances are $3.2\pm0.5$ kpc  for 4U 0614+091,  $5.4\pm0.8$ kpc for 2S 0918-549 and $4.3\pm0.6$ kpc for 4U 1246-588.}\\
{$^b$ The bolometric correction factor are estimated by the best-fitting model and assuming the power-law spectral component up to around 100 keV and down to around 0.1 keV. }\\
{$^c$ The bolometric luminosities are in the energy band of 0.1--100 keV.
The Eddington luminosity is calculated by assuming that the neutron star's mass is $1.4~{\rm M}_{\odot}$. Therefore, the Eddington luminosity is $2.0\times10^{38}{\rm ~erg~s^{-1}}$ for hydrogen-rich photosphere and $3.5\times10^{38}{\rm ~erg~s^{-1}}$ for hydrogen-poor photosphere(\citealt{Zand+etal+2007}). 
}

}
\end{tablenotes}
\label{table2}
\end{sidewaystable*}

\subsection{2S 0918-549}
The joint X-ray spectra of 2S 0918-549 with the Swift/XRT and RXTE/PCA observations at around MJD 55220.4 was modeled in the energy range of 0.6--20 keV. Similar to the spectral analysis for 4U 0614+091, the spectral model consisted of a power-law spectral component, a blackbody component and a Gaussian line emission at around 0.6 keV.  We fixed the Gaussian emission line to 0.635 keV (according to \citealt{Juett+etal+2001}). We obtained the best-fitting parameters and derived the hard-to-soft transition flux in 0.6--20 keV as $(2.4^{+0.2}_{-0.1})\times10^{-10}$ erg s$^{-1}$ cm$^{-2}$, corresponding to an X-ray luminosity of $(8.4\pm 2.5) \times10^{35}$ erg s$^{-1}$ by taking the source distance of  $5.4\pm0.8$ kpc. By the bolometric correction factor of  $1.96^{+0.07}_{-0.04}$ derived from the spectral parameters, the mass accretion rate corresponds to $0.0082\pm0.0035~\dot{\rm M}_{\rm Edd}$ for a hydrogen-rich photosphere.

The PCA spectrum in the energy of 3--20 keV at around MJD 55202.7 was fitted with only an absorbed power-law model with a fixed $N_{\rm H}$ of $3.0\times10^{21}$ cm$^{-2}$, since 2S 0918-549 was in the hard state and the low count rate in the PCA data did not allow us to include more spectral components. The measured X-ray flux in 3--20 keV was $(8.7\pm0.2)\times10^{-11}$ erg s$^{-1}$ cm$^{-2}$ and the corresponding X-ray luminosity is  $(3.0\pm 0.9)\times10^{35}$ erg s$^{-1}$. From the fitting spectral parameters, we estimated the bolometric correction factor as $4.6^{+0.5}_{-0.4}$. Therefore, the mass accretion rate corresponding to the hard-to-soft transition could be estimated as  $0.0070\pm0.0030~\dot{\rm M}_{\rm Edd}$ for a hydrogen-rich photosphere. 

Similarly, from observations at MJD 55214.9 and MJD 55240.5, we obtained the soft-to-hard transition fluxes (in table~\ref{table2}) and inferred the mass accretion rates as $0.011\pm0.005~\dot{\rm M}_{\rm Edd}$ and $0.0095\pm0.0040~\dot{\rm M}_{\rm Edd}$ for a hydrogen-rich photosphere, respectively. The mass accretion rate corresponding to soft-to-hard transition is consistent with those corresponding to hard-to-soft transition.  

\subsection{4U 1246-588}
For 4U 1246-588, there were only RXTE observations covering the period when the spectral state transitions occurred. We modeled the corresponding PCA spectra on MJD 55573.6, MJD 55645.1, MJD 55701.6 and MJD 55865.9 in the energy range between 3 and 20 keV, with an absorbed power-law, and a fixed $N_{\rm H}$ of $2.5\times10^{21}$ cm$^{-2} $(\citealt{Zand+etal+2008}). The lowest flux measured from the four observations was  $(5.6\pm0.5)\times10^{-11}$ erg s$^{-1}$ cm$^{-2}$.  Therefore, the possible lower limit on the hard-to-soft transition luminosity is $(1.2\pm0.4)\times10^{35}$ erg s$^{-1}$. From the fitting spectral parameters, we estimated that the corresponding mass accretion rate is $0.0022^{+0.0012}_{-0.0010}~\dot{\rm M}_{\rm Edd}$ by taking the bolometric correction factor as  $3.6^{+0.9}_{-0.3}$  and a hydrogen-rich photosphere. This gives the lowest mass accretion rate among the three sources, which possibly represents the lower limit of X-ray luminosity of the hard-to-soft transition in all LMXBs. Due to the less intensive coverage of the spectral evolution of this source in the observations we analysed, the value of the lower limit might suffer from the sparse sampling of the spectral evolution.

On the other hand, we also obtained the soft-to-hard transition fluxes in the other four observations and inferred the lowest accretion rate as $0.0028^{+0.0013}_{-0.0012}~\dot{\rm M}_{\rm Edd}$ for a hydrogen-rich photosphere, which is consistent with the accretion rate corresponding to hard-to-soft transition.

%________________________________________________Discussion
\section{Discussion}
\subsection{The lowest luminosity of the hard-to-soft spectral state transitions in NS LMXBs}
\cite{Yu+etal+2004,Yu+etal+2007} have discovered the correlation between the luminosity corresponding to the hard-to-soft transition and the peak luminosity of the following soft state in some black hole and neutron star LMXBs. \cite{Yu+Yan+2009} have demonstrated that such a correlation applies to all the bright Galactic X-ray binaries, including both transient and persistent sources. In the current comprehensive study of the X-ray data from those all-sky-monitors covering soft and hard X-ray components, the luminosity corresponding to the hard-to-soft state transitions in UCXBs, such as those in 2S 0918-549 and 4U 0614+091, are among the lowest luminosity in unit of Eddington in the Galactic binary sample (see \citealt{Yu+Yan+2009}). Therefore accurate measurements of the transition luminosity in the ultra-compact X-ray binaries would set the lower limit of the hard-to-soft transition in Galactic X-ray binaries.  In the current study, we have analysed the Swift and RTXE observations of the hard-to-soft spectral state transitions in a few ultra-compact NS LMXBs which indeed represent the lowest hard-to-soft spectral state transitions in the Galactic X-ray binaries (see \citealt{Yu+Yan+2009}).  The lowest luminosity corresponding to the hard-to-soft transition is found in the range ${\rm 0.002-0.04}~{\rm L}_{\rm Edd}$ (assuming a hydrogen-rich photosphere) or ${\rm 0.001-0.02}~{\rm L}_{\rm Edd}$(assuming a hydrogen-poor photosphere).

The fact that the ultra-compact X-ray binaries displays the lowest hard-to-soft transition luminosity is not by accident. We know that their disc mass is small due to the size of their Roche lobe and relatively low level of accretion (see also discussion on disc mass in \citealt{Yu+Dolence+2007,Yu+etal+2007}). \cite{Wu+etal+2010} have shown that in the outbursts of soft X-ray transients, there is a correlation between the peak luminosity of an outburst and the orbital period.  The correlation suggests that short-period LMXBs, such as UCXBs, usually have weaker and shorter outbursts. On the other hand, \cite{Zand+etal+2007} have also found the mass-transfer rate of almost all persistent UCXBs are less than $0.02~\dot{\rm M}_{\rm Edd}$ except 4U 1820-30, by assuming that the average accretion rate is roughly equal to the mass-transfer rate in these persistent sources. From the above studies, the luminosity of the hard-to-soft transition is found in the range between 0.002 and 0.04$~{\rm L}_{\rm Edd}$, yielding the lowest luminosity limit of the hard-to-soft transition in LMXBs. As the most extreme case,  4U 1246-588 contributes to the measurement of the lowest hard-to-soft transition luminosity among the current ultracompact samples, corresponding to the mass accretion rate level around $0.002~\dot{\rm M}_{\rm Edd}$. It is worth noting that, some known or suspected UCXBs  (e.g. 1RXS J172525.2-325717 and NGC 6440 X-2 ) can probably contribute even lower state transition luminosity, but it's difficult to locate any state transition and obtain exact transition flux to these sources based with current observation data.

\subsection{Critical mass accretion rate: No significant difference between BH and NS binaries}
In the results of \cite{Yu+Yan+2009} and \cite{Tang+etal+2011} which are based on X-ray monitoring observations over more than a decade,  BH HMXBs such as Cygnus X-1 and persistent ultra-compact NS LMXBs have the lowest luminosity of the hard-to-soft transition among the bright Galactic X-ray binaries. Among the ultracompact NS LMXBs, 4U 1820-30 and IGRJ 17473-2721 are representative sources with the highest hard-to-soft transition luminosity at about $\sim0.07~L_{\rm Edd}$ in 15--50 keV in all neutron star LMXBs. Here we look at the low luminosity end. We have determined that the lowest luminosity of the hard-to-soft and soft-to-hard transition in NS LMXBs is in the ranges of ${\rm 0.002-0.04}~{\rm L}_{\rm Edd}$ and ${\rm 0.003-0.05}~{\rm L}_{\rm Edd}$, respectively.

Considering the transition of the accretion regimes from low mass accretion rates to high mass accretion rates, corresponding to the case during the rising phase of the outbursts in BH and NS LMXB transients, there should exist substantial differences between NS systems and BH systems in accretion theory.  The most representative model is the Advection Dominated Accretion Flow (ADAF) model (\citealt{Narayan+Yi+1995,Esin+etal+1997}). \cite{Narayan+Yi+1995} have derived a critical accretion rate $M_{\rm crit}\sim0.1\alpha^2{\rm M_Edd}$ for neutron star LMXBs in the ADAF model, in which the critical accretion rate for the BH systems is found an order of magnitude higher if we assume the viscosity coefficient $\alpha$ is the same in BH systems and NS systems. This allows us to test if the ADAF model applies to both BH and NS systems or the ADAF model is not applicable to NS LMXBs.  

The black hole binary Cygnus X-1 is the reference BH X-ray binary which displays the lowest transition luminosity (hard-to-soft transition) in the BHBs  (see \citealt{Yu+Yan+2009}). Its bolometric luminosity corresponding to the X-ray spectral state transitions was nearly constant  in a range of a few tens percent and was measured at around $4~\times{10}^{37}$ erg s$^{-1} $ by assuming the distance as 2.5 kpc (\citealt{Zhang+etal+1997}). However the most recent measurement from a trigonometric parallax suggests the distance of Cygnus X-1 is 1.86 kpc (\citealt{Reid+etal+2011}). The luminosity can be re-estimated as $2.2~\times{10}^{37}$ erg s$^{-1} $, corresponding to $0.01~{\rm L}_{\rm Edd}$ if the black hole mass in Cygnus X-1 is $15~{\rm M}_{\odot} $(\citealt{Orosz+etal+2011}). And its luminosity corresponding to the soft-to-hard transition is close to that corresponding to the hard-to-soft transition (\citealt{Zhang+etal+1996}). The mini-outbursts in black hole binary GRS 1739-278 even showed lower bolometric luminosities, which are $8.5~\times{10}^{36}$ erg s$^{-1}$ for the hard-to-soft transition and $2.2~\times{10}^{36}$ erg s$^{-1}$ for the soft-to-hard transition if the bolometric correction factor is taken as 2 for 0.5--10 keV energy band when the distance is taken as 7.5 kpc (\citealt{Yan+Yu+2017}). Notice that we here use the data of the last hard state before the state transition for GRS 1739-278.  They correspond to $0.007~{\rm L}_{\rm Edd}$ and $0.002~{\rm L}_{\rm Edd}$ respectively, by assuming the BH mass as $8~{\rm M_\odot}$. Here we consider bolometric luminosity, but notice that the energy band in figure 6 of \cite{Yan+Yu+2017} was 15--50 keV for Cygnus X-1 and sources other than GRS 1739-278. Based on the radiative efficiency model from \cite{Hopkins+etal+2006}, the lowest critical mass accretion rate in BHBs can then be estimated as $0.009-0.01~\dot{\rm M}_{\rm Edd}$ for the hard-to-soft transition and $0.004-0.01~\dot{\rm M}_{\rm Edd}$ for the soft-to-hard transition.

The hard-to-soft and the soft-to-hard transition luminosity we measured in the three persistent ultracompact NS LMXBs is consistent with that observed in Cygnus X-1 and GRS 1739-278. Note that, if the accreted material is hydrogen-poor, which is likely case in UCXBs, the mass accretion rate is in the range of ${\rm 0.001-0.02}~\dot{\rm M}_{\rm Edd}$ for hard-to-soft transition and  ${\rm 0.002-0.03}~\dot{\rm M}_{\rm Edd}$ for soft-to-hard transition, which are still comparable with the mass accretion rate in Cygnus X-1 and GRS 1739-278. It seems that the ADAF model is not consistent with our results and the accretion regimes, as we observed, may be not of the ADAF or not set by the ADAF thresholds at all. However, the ADAF model would still apply to NS systems if the viscosity coefficient $\alpha$ is a few times higher than that in BH systems. Another possible scenario is that the viscosity coefficient $\alpha$ is significantly larger in hydrogen-poor accreted material, but there is no evidence to support this scenario yet (\citealt{Lasota+etal+2008}). In summary of the observed results, we found no apparent difference in the transition luminosity in BH and NS systems.

\subsection{No hysteresis effect in persistent UCXBs}

As shown in table~\ref{table2}, there is no hysteresis in the three UCXB samples, 4U 0614+091, 2S 0918-549 and 4U 1246-588. Even surprising, their soft-to-hard luminosity is slightly higher than the hard-to-soft luminosity. The hysteresis effect seems to associate with large amplitude outbursts (\citealt{Miyamoto+etal+1995,Maccarone+Coppi+2003,Yu+etal+2003,Gladstone+etal+2007}).  For persistent sources,  the hysteresis effect is relatively weak but still can be found in some persistent NS LMXBs, such as 4U 1636-53, 4U 1705-44 and  4U 1728-34 (\citealt{Darias+etal+2014,Gladstone+etal+2007}). Hysteresis effect seems to disappear in the persistent UCXBs.  The hysteresis effect of spectral state transitions is possibly related to the stable disc mass in the systems as discussed in the following subsection.

\subsection{Low transition luminosity and less mass in the disc: similarity between ultracompact LMXBs and HMXBs ?}
The systematic study of the hard-to-soft spectral state transitions in Galactic X-ray binaries indicate that the transition luminosity in Eddington units varies in a range spanning by more than two order of magnitude (\citealt{Yu+Yan+2009}).  A possible explanation is that the instantaneous mass accretion rate is not the only parameter which determines the luminosity at which the state transition occurs (\citealt{Miyamoto+etal+1995,Homan+etal+2001}).  From observational perspective, we need to find dominant physical parameters which determines spectral state transitions other than the mass accretion rate. Quit a few observations have indicated that disc mass is apparently another dominant parameter (see \citealt{Yu+etal+2004,Yu+etal+2007}). In the case of the outbursts in the black hole transient GX 339-4, there is a correlation between the outburst peak flux and the outburst waiting time, while the luminosity of the hard-to-soft transition is correlated with the outburst peak flux. The mass in disc involved in an outburst should be correlated with the total radiative energy released during the entire outburst.  A longer waiting time for an outburst would suggest that a more massive disc is involved in the outburst if the mass transfer rate from the companion star is nearly constant.  

Whether disc mass or disc size plays the  role in determination of the spectral transitions can be investigated in the observations of spectral state transitions in ultracompact LMXBs and HMXBs. Contrary to the three ultracompact LMXBs studied in the current study, in the ultra-compact NS LMXB 4U 1820-30, the transition luminosity is about two orders of magnitude higher. However the orbital period of 4U 1820-30 is only 11 minutes, which means its disc size is about the smallest among all known UCXBs (\citealt{Zurek+etal+2009}). 
Similarly, in the study presented in \cite{Yu+Yan+2009}, Cygnus X-1 has a much lower transition luminosity than that of another HMXB Cygnus X-3, which means that the transition luminosity still spans in a large luminosity range in HMXBs.
Therefore, disc size, which limited by the Roche lobe radii,  is not the other parameter which determines spectral transitions. 

 The disc instability model (DIM) suggests that these is a steady-state accretion disc in persistent LMXBs, because the mass transfer rate is always higher than the critical value (\citealt{Lasota+etal+2008,Zand+etal+2007}). For standard thin disc, the surface density  is positively correlated with the mass accretion rate.  Since 4U 1820-30 usually has a much higher mass accretion rate,  its the disc mass is higher than those of other UCXBs although its disc size is smaller. This supports the mass in the accretion disc determines the spectral transition rather than the disc size. 

%________________________________________________Conclusion
\section{Conclusion}
We have determined the spectral state transitions with the lowest X-ray luminosity in the known or candidate UCXBs sample based on X-ray monitoring observations with the RXTE/ASM, the Swift/BAT and the MAXI.  These state transitions were identified in three sources, namely 4U 0614+091, 2S 0914-549 and 4U 1246-588. We then measured the X-ray flux corresponding to the hard-to-soft and the soft-to-hard transitions by  analysing pointed observations with the Swift/XRT and RXTE/PCA in the public archive.  We then derived the X-ray flux corresponding to the state transitions in these neutron star LMXBs. We conclude

\begin{itemize}

\item{
We have found that 4U 1246-588, 4U 0614+091, and 2S 0918-549 show the lowest luminosity of the hard-to-soft transition luminosity in all the X-ray binaries (see \citealt{Yu+Yan+2009,Tang+etal+2011}). The mass accretion rates corresponding to the hard-to-soft and the soft-to-hard transitions at the lowest luminosities are in the range $0.002-0.04~\dot{\rm M}_{\rm Edd}$ and $0.003-0.05~\dot{\rm M}_{\rm Edd}$, respectively, where the Eddington luminosity is taken as $2.0\times10^{38}{\rm ~erg~s^{-1}}$ and the radiative efficiency is taken as 0.1. 

}

\item{
We found no hysteresis effect of spectral state transitions in 4U 0614+091, 2S 0918-549 and 4U 1246-588, which is similar to the HMXB Cygnus X-1.  This is possibly related to the stable disc mass in these systems during an accretion flare when both the hard-to-soft and the soft-to-hard spectral state transition occurs.  We found that the lowest accretion rate corresponding to the state transition in UCXBs is comparable to that of the Cygnus X-1 and GRS 1739-278, which indicates that the state transition luminosity is not affected by the nature of the compact stars and whether the compact star has a hard surface or not.  Our results suggest that either ADAF model does not apply in the hard state of black hole and neutron star X-ray binaries or the thresholds of critical mass accretion rate in ADAF do not apply in state transitions.
}

\item{
The general long-term behaviour of the X-ray flux vs power-law index of UCXBs looks similar to those in HMXBs, probably because they both have less massive accretion discs comparing with other LMXBs.  The similarity in these the low luminosity of the X-ray spectral state transitions in UCXBs and HMXBs seem to support that disc mass is taking a dominant role rather the disc size in affecting the transition and producing hysteresis effect, although the detailed mechanism is not clear. 
}

\end{itemize}

%________________________________________________Acknowledgments
\section*{Acknowledgments}
We would like to thank the RXTE and the Swift Guest Observer Facilities at NASA Goddard Space Flight Center for providing the RXTE/PCA and Swift/XRT products and the Swift/BAT and RXTE/ASM transient monitoring results. This research has made use of MAXI data provided by RIKEN, JAXA and the MAXI team.  We appreciate stimulating comments and suggestions by the anonymous referee which have improved the manuscript significantly.  W.Y. would like to acknowledge the support by the National Program on Key Research and Development Project (Grant No. 2016YFA0400804), by the National Natural Science Foundation of China under grant No. 11333005 and by the FAST fellowship, which is supported by Special Funding for Advanced Users, budgeted and administrated by Center for Astronomical Mega-Science, Chinese Academy of Sciences (CAMS).
\nocite{*}

%________________________________________________References
%\bibliographystyle{mn2e}
%\bibliography{references}

\clearpage

\label{lastpage}

\end{document}